\documentclass[12pt]{article}
\usepackage{rotating}
\usepackage{amssymb,amsmath,amsthm,amscd,latexsym}
\usepackage{amsfonts}
\usepackage{tikz}
\usepackage{rotating}
\usetikzlibrary{arrows,positioning,shapes,fit,calc}
\usepackage{tikz-cd}
\usepackage[mathscr]{eucal}
\usepackage{color,hyperref}
\usepackage{graphicx}
\usepackage{longtable}
\usepackage{multirow}
\usepackage[ruled,vlined]{algorithm2e}

 \theoremstyle{definition}
 
 \theoremstyle{remark}

 \numberwithin{equation}{section}
\begin{document}

\title{New Extremal Binary Self-Dual Codes of Length 72 from $M_6(\mathbb{F}_2)G$ - Group Matrix Rings by a Hybrid Search Technique Based on a Neighbourhood-Virus Optimisation Algorithm }
\author{ Adrian Korban \\
Department of Mathematical and Physical Sciences \\
University of Chester\\
Thornton Science Park, Pool Ln, Chester CH2 4NU, England \\
Serap \c{S}ahinkaya \\
Tarsus University, Faculty of Engineering \\ Department of Natural and Mathematical Sciences \\
Mersin, Turkey \\
Deniz Ustun \\
Tarsus University, Faculty of Engineering \\ Department of Computer Engineering \\
Mersin, Turkey}
 \maketitle

\begin{abstract}
In this paper, a new search technique based on the virus optimisation algorithm  is proposed for calculating the neighbours of binary self-dual codes. The aim of this new technique is to calculate neighbours of self-dual codes without reducing the search field in the search process (this is a known in the literature approach due to the computational time constraint) but still obtaining results in a reasonable time (significantly faster when compared to the standard linear computational search). We employ this new search algorithm to the well-known neighbour method and its extension, the $k^{th}$-range neighbours and search for binary $[72,36,12]$ self-dual codes. In particular, we present six generator matrices of the form $[I_{36} \ | \ \tau_6(v)],$ where $I_{36}$ is the $36 \times 36$ identity matrix, $v$ is an element in the group matrix ring $M_6(\mathbb{F}_2)G$ and $G$ is a finite group of order 6, which we then employ to the proposed algorithm and search for binary $[72,36,12]$ self-dual codes directly over the finite field $\mathbb{F}_2$. We construct 1471 new Type I binary $[72, 36, 12]$ self-dual codes with the rare parameters $\gamma=11, 13, 14, 15, 17, 19, 20, 21, 22, 23, 25, 26, 28, 29, 30, 31, 32$ in their weight enumerators.
\end{abstract}


\section{Introduction}\label{intro}
Self-dual codes over finite fields are a class of linear block codes that have been extensively studied in recent years. A very well known  technique for producing extremal binary self-dual codes is to consider generator matrices of the form $[I_n \ | \ A_n],$ where $I_n$ is the $n \times n$ identity matrix and $A_n$ is some $n \times n$ matrix with entries from a finite field $\mathbb{F}_2.$ In \cite{Gildea1}, the authors consider a special type of the matrices $A_n$ to reduce the search field. They employ group rings and a well-established map, $\sigma ,$ that sends a group ring element $v$ to some $n \times n$ matrix that is fully defined by the elements appearing in the first row - these elements are from the ring $R$ or a finite field $\mathbb{F}_q.$  One can see \cite{Dougherty01, Dougherty02, Dougherty1,Dougherty3,KSD1} for examples of this technique. In \cite{Dougherty2}, the authors employ another map, $\Omega(v) ,$ that sends group ring elements to some more complex $n \times n$ matrices, called composite matrices, that are fully defined by the elements in the first row. They consider generator matrices of the form $[I_n \ | \ \Omega(v)]$, to obtain some binary self-dual codes with. More examples of this approach can be found in \cite{Dougherty2, Dougherty4, Dougherty5}. Recently in \cite{Dougherty6}, the authors extend the map $\sigma$ and consider elements from the group matrix ring $M_k(R)G$ rather than elements from the group ring $RG.$ They define a map that sends an element from the group matrix ring $M_k(R)G$ to a $kn \times kn$ matrix over the ring $R.$ They call this map $\tau_k(v).$ The motivation of this extension is to use generator matrices of the form $[I_{kn} \ | \ \tau_k(v)]$ to obtain codes with parameters that could not be obtained from the generator matrices of the forms $[I_n \ | \ \sigma(v)]$ or $[I_n \ | \ \Omega(v)].$ Please see \cite{Dougherty3} for more details. Another technique for obtaining new binary self-dual codes is to consider the well-known neighbour method.
Two binary self-dual codes of length $%
2n$ are said to be neighbours of each other if their intersection has
dimension $n-1$. Let $\mathcal{C}$ be a self-dual code of length $2n$ and let $\mathbf{x}\in {\mathbb{F}}_{2}^{2n} \setminus \mathcal{C},$ then
$\mathcal{D}=\left\langle \left\langle \mathbf{x} \right\rangle ^{\bot }\cap \mathcal{C%
},\mathbf{x} \right\rangle $ is a neighbour of $\mathcal{C}$ -- from this definition it follows that one needs to find a vector $\mathbf{x}$ of length $2n$ that satisfies the given conditions to obtain a neighbour of the code $\mathcal{C}.$ This is, computationally, a difficult task as the search field can get very big for codes of great lengths, for instance, to find the neighbours of a binary self-dual code of length 64, one would need to consider $2^{64}$ possibilities to find the right vectors. To overcome this problem to some degree, that is, to reduce the search field, some researchers fix some of the entries of the vector $\mathbf{x}$ when calculating the possible neighbours. This approach can be found in \cite{Dougherty1, Dougherty2, Dougherty4, Dougherty5}, where the authors find neighbours of codes of length 68 by fixing the first 34 entries of the vector $\mathbf{x}$ to be zeros. By reducing the search field we are automatically reducing the number of possible neighbours which could have parameters in their weight enumerators that were not known before. Typically, by reducing the search field for the vector $\mathbf{x},$ the neighbours obtained have parameters within a narrow interval. In \cite{GN}, the above definition of a neighbour is extended to the $k^{th}$-range neighbours, i.e., let $\mathbf{x}_0 \in \mathbb{F}_2^{2n} \setminus \mathcal{N}_{(0)},$ then

$$\mathcal{N}_{(i+1)}=\left\langle \left\langle \mathbf{x}_i \right\rangle^{\bot} \cap \mathcal{N}_{(i)}, \mathbf{x}_i \right\rangle,$$
where $\mathcal{N}_{(i+1)}$ is the neighbour of $\mathcal{N}_{(i)}$ and $\mathbf{x}_i \in \mathbb{F}_2^{2n} \setminus \mathcal{N}_{(i)}.$ This method simply involves calculating the neighbours of a self-dual code $\mathcal{N}_{(0)}$ and then taking at random one of its neighbours, $\mathcal{N}_i,$ and calculating the possible neighbours of $\mathcal{N}_i$ and repeating the process over again. The computational constraint here is that one may end up getting many neighbours of the binary self-dual code $\mathcal{N}_{(i)}.$ One then has to consider their neighbours one at a time which can be impractical.

In this work, a novel hybrid search technique called the neighbourhood–virus optimisation algorithm is proposed for calculating the neighbours of binary self-dual codes. In this technique, we employ a virus optimisation algorithm since such algorithms have proven to be extremely effective tools for problems with big size search fields, please for example see \cite{KSD1, KSD2, KSD3, KSD4}. Our new search technique allows one to calculate the neighbours of binary self-dual codes without reducing the search field for the vector $\mathbf{x}$ and also, the new technique allows one to calculate the $k^{th}$-range neighbours of codes with parameters within a pre-selected range in their weight enumerators without going over the neighbours one at a time which can be time consuming. We particularly focus on calculating neighbours of binary $[72,36,12]$ self-dual codes since there are still many codes of this type with unknown parameters in their weight enumerators. For this reason, we employ some generator matrices of the form $[I_{kn} \ | \ \tau_{kn}]$ with the group matrix ring $M_6(\mathbb{F}_2)G,$ where $G$ is a group of order 6, to first search for binary $[72,36,12]$ self-dual codes and we next employ our new neighbourhood-virus algorithm to calculate their possible neighbours. In this way, we obtain many new Type I binary $[72,36,12]$ self-dual codes with various parameters in their weight enumerator that were not known in the literature before.

The motivations for introducing the new search algorithm are:
\begin{itemize}
  \item The neighbour method is an effective tool to search for new extremal binary self-dual codes, please see \cite{Dougherty5,Gulliver1, KSD4} for some examples.
  \item The virus optimisation algorithm (VOA), one of the well-known optimisation algorithms, is known for coping with large search fields significantly better and finds codes much faster than the standard linear search, please see \cite{KSD3, KSD4} for more details in this direction.
\end{itemize}

The novelties of the proposed algorithm are:
\begin{itemize}
  \item   One can calculate the $k^{th}$-range neighbours of codes with parameters within a pre-selected range in their weight enumerators - this, as we show by our computational results, can lead to obtaining many new binary self-dual codes of a particular length.
  \item   One does need to reduce the search field of the vector $\mathbf{x};$ as mentioned earlier, typically to overcome the computational time constraint, the first half of the entries of the vector $\mathbf{x}$ are set to be all zero, and the rest of entries are chosen randomly, while in our search algorithm, all of entries of the vector $\mathbf{x}$ are chosen by the virus optimisation algorithm in the hybrid search scheme.
\end{itemize}

The rest of the work is organised as follows.  In Section~2, we give preliminary definitions and results on self-dual codes, special matrices, group rings and  recall the map $\tau_k(v)$.
In Section~3, we introduce the new search algorithm and explain how to employ it to the well-known neighbour method and its extension, the $k^{th}$-range neighbours.
In Section~4, we present 6 generator matrices of the form $[I_{36} \ | \ \tau_6(v)]$ where for each generator matrix, we fix the $6 \times 6$ matrices by letting them be either circulant and reverse circulant matrices. We then use these generator matrices to search for binary $[72,36,12]$ self-dual codes.
We next use our new search algorithm employed to the well-known neighbour method and its extension to search for possible neighbours of the earlier obtained binary self-dual codes of length 72. As a result, we find 1471 Type I binary $[72,36,12]$ self-dual codes with parameters in their weight enumerators that were not previously known. We only give the new parameters $[\gamma, \beta]$ in order to save space. The vector $\mathbf{x}$, the generator matrix, the automorphism group, and the partial weight distribution of each code are available online at \cite{sahinkaya}. We finish with concluding remarks and directions for possible future research.

\section{Preliminaries}

In this section we recall some well-known definitions and notions on codes, special matrices, group rings and the known map $\tau_{kn}.$

A code $\mathcal{C}$ of length $n$ over a Frobenius ring $R$ is a subset of $R^n$. If the code is
a submodule of $R^n$ then we say that the code is linear. The minimum distance $d$ of a linear code $\mathcal{C}$ is determined by:

$$d= min_{\mathbf{x} \neq \mathbf{y}}d_H(\mathbf{x}, \mathbf{y}),$$

where $d_H(\mathbf{x}, \mathbf{y})$ denotes the Hamming distance between codewords $\mathbf{x}, \mathbf{y} \in C$.
Hamming distance is defined as the number of positions that differs between two distinct codewords. Let $\mathbf{x}=(x_1,x_2,\dots,x_n)$
and $\mathbf{y}=(y_1,y_2,\dots,y_n)$ be two elements of $R^n.$ Then
\begin{equation*}
\langle \mathbf{x},\mathbf{y} \rangle_E=\sum x_iy_i.
\end{equation*}
The dual $\mathcal{C}^{\bot}$ of the code $\mathcal{C}$ is defined as
\begin{equation*}
\mathcal{C}^{\bot}=\{\mathbf{x} \in R^n \ | \ \langle \mathbf{x},\mathbf{y}
\rangle_E=0 \ \text{for all} \ \mathbf{y} \in \mathcal{C}\}.
\end{equation*}
We say that $\mathcal{C}$ is self-orthogonal if $\mathcal{C} \subseteq \mathcal{C}^\perp$ and is self-dual
if $\mathcal{C}=\mathcal{C}^{\bot}.$

An upper bound on the minimum Hamming distance of a binary self-dual code
was given in \cite{Rains1}. Let $d_{I}(n)$ and $d_{II}(n)$ be the
minimum distance of a Type~I (singly-even) and Type~II (doubly-even) binary code of length $n$,
respectively. Then
\begin{equation*}
d_{II}(n) \leq 4\lfloor \frac{n}{24} \rfloor+4
\end{equation*}
and
\begin{equation*}
d_{I}(n)\leq
\begin{cases}
\begin{matrix}
4\lfloor \frac{n}{24} \rfloor+4 \ \ \ if \ n \not\equiv 22 \pmod{24} \\
4\lfloor \frac{n}{24} \rfloor+6 \ \ \ if \ n \equiv 22 \pmod{24}.%
\end{matrix}%
\end{cases}%
\end{equation*}

Self-dual codes meeting these bounds are called \textsl{extremal}.

In this work, we find many extremal binary self-dual codes with parameters $[72,36,12].$ The possible weight enumerators for a Type~I $[72,36,12]$ codes are as follows (\cite{Dougherty1}):
$$W_{72,1}=1+2\beta y^{12}+(8640-64\gamma)y^{14}+(124281-24\beta+384\gamma)y^{16}+\dots$$
$$W_{72,2}=1+2 \beta y^{12}+(7616-64 \gamma)y^{14}+(134521-24 \beta+384 \gamma)y^{16}+\dots$$
where $\beta$ and $\gamma$ are parameters.
The possible weight enumerators for Type~II $[72,36,12]$ codes are (\cite{Dougherty1}):
$$1+(4398+\alpha)y^{12}+(197073-12\alpha)y^{16}+(18396972+66\alpha)y^{20}+\dots $$
where $\alpha$ is a parameter.
For an up-to-date list of all known Type~I and Type~II binary self-dual codes with parameters $[72,36,12]$ please see \cite{selfdual72}.

A circulant matrix is one where each row is shifted one element to the right relative to the preceding row and a reverse circulant
matrix is one where each row is shifted one element to the left relative to the preceding row.
We label the circulant matrix as $A=circ(\alpha_1,\alpha_2\dots , \alpha_n),$ and the reverse circulant matrix as $A=revcirc(\alpha_1,\alpha_2\dots , \alpha_n),$  where $\alpha_i$ are ring elements.
The transpose of a matrix $A,$ denoted by $A^T,$ is a matrix whose rows are the columns of $A,$ i.e., $A^T_{ij}=A_{ji}.$ A block  circulant matrix is one where each block  is shifted one block  to the right relative to the preceding block and a block  reverse circulant
matrix is one where each block is shifted one block to the left relative to the preceding block.
We label the circulant matrix as $A=CIRC(A_1, A_2\dots , A_n),$ and the reverse circulant matrix as $A=REVCIRC(A_1, A_2\dots , A_n),$  where $A_i$ are $k \times k$ matrices.

Let $G$ be a finite group of order $n$, then the group ring $RG$
consists of $\sum_{i=1}^n \alpha_i g_i$, $\alpha_i \in R$, $g_i \in G.$

Addition in the group ring is done by coordinate addition, namely
\begin{equation}
\sum_{i=1}^n \alpha_i g_i +\sum_{i=1}^n \beta_i g_i =\sum_{i=1}^n (\alpha_i
+ \beta_i)g_i.
\end{equation}
The product of two elements in a group ring is given by
\begin{equation}
\left(\sum_{i=1}^n \alpha_i g_i \right)\left(\sum_{j=1}^n \beta_j g_j
\right)= \sum_{i,j} \alpha_i \beta_j g_i g_j.
\end{equation}
It follows that the coefficient of $g_k$ in the product is $\sum_{g_i
g_j=g_k} \alpha_i \beta_j.$

We now recall the map $\tau_k(v),$ where $v \in M_k(R)G$ and where $M_k(R)$ is a non-commutative Frobenius matrix ring and $G$ is a finite group of order $n,$ that was introduced in \cite{Dougherty2}.

Let $v=A_{g_1}g_1+A_{g_2}g_2+\dots+A_{g_n}g_n \in M_k(R)G,$ that is, each $A_{g_i}$ is a $k \times k$ matrix with entries from the ring $R.$ Define the block matrix $\tau_k(v) \in (M_{k}(R))_n$ to be

\begin{equation}\label{sigmakv}
\tau_k(v)=
\begin{pmatrix}
A_{g_1^{-1}g_1} & A_{g_1^{-1}g_2} & A_{g_1^{-1}g_3} & \dots &
A_{g_1^{-1}g_{n}} \\
A_{g_2^{-1}g_1} & A_{g_2^{-1}g_2} & A_{g_2^{-1}g_3} & \dots &
A_{g_2^{-1}g_{n}} \\
\vdots & \vdots & \vdots & \vdots & \vdots \\
A_{g_{n}^{-1}g_1} & A_{g_{n}^{-1}g_2} & A_{g_{n}^{-1}g_3} & \dots &
A_{g_{n}^{-1}g_{n}}
\end{pmatrix}
.
\end{equation}

For a given element $v \in M_k(R)G,$ we construct the matrix $\tau_k(v)$ by viewing each element in a $k$ by $k$ matrix as an element in the larger matrix.
\begin{equation}
B_k(v)=\langle \tau_k(v) \rangle.
\end{equation}
Here the code $B_k(v)$ is formed by taking all linear combinations of the rows of the matrix with coefficients in $R$. In this case the ring over which the code is defined is commutative so it is both a left linear and right linear code.

\section{The Neighbourhood-Virus Optimisation Algorithm}

The virus optimisation algorithm (VOA) (\cite{Cuevas}) is one of the heuristic optimisation methods and it has been recently used for the problem of finding self-dual codes in \cite{KSD3, KSD4}. The VOA iteratively optimises complex and hard systems or problems by using the population-based search techniques. The VOA mimics the behaviours of viruses to a living cell and has a replication step that is performed by using viruses that are classified into common and strong - this classification depends on some properties of the vectors, in our case, it will be the hamming weight. There is also an immune system step in VOA and this step is used for coping with the uncontrolled increase of viruses in the population. In the algorithm, the aim of separating viruses in the population is to provide a balance between the exploitation and exploration search abilities of VOA.

We now describe the new search algorithm and how it is employed to search for neighbours of binary $[72,36,12]$ self-dual codes with new $[\gamma, \beta]$ parameters in their weight enumerators. The described approach can be easily adapted to search for neighbours of binary self-dual codes of other lengths.

\begin{sidewaysfigure}
\centering
\includegraphics[width=220mm, height=90mm]{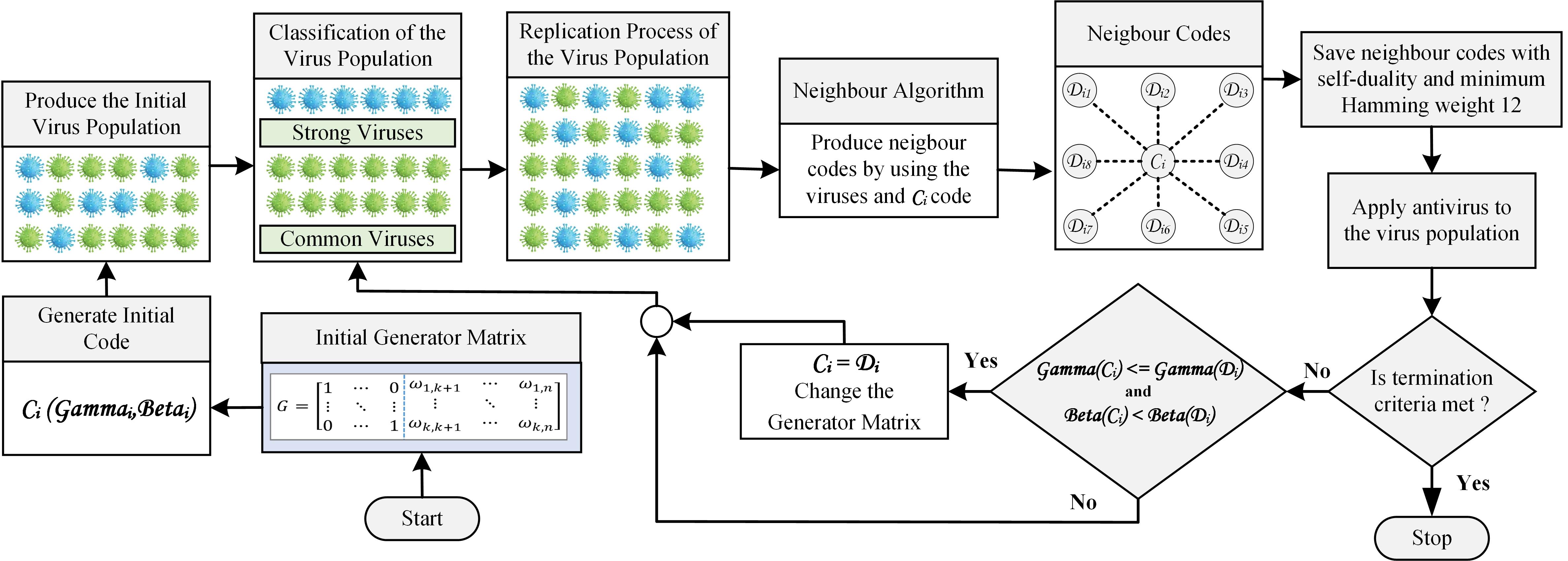}
\caption{Flowchart of the Neighbourhood-Virus Optimisation Algorithm}\label{VOA}
\end{sidewaysfigure}

In the proposed hybrid search technique, we first construct a binary self-dual code from a generator matrix that we refer to as an initial generator matrix. At this very first step, we can employ different generator matrices that produce self-dual codes. Next, the initial virus population is randomly produced. Each virus in the population is represented by a vector $\mathbf{x}$. After the codes are produced by using the well-known neighbour method, the hamming weights of these codes are computed. The viruses are classified into strong and common viruses according to their minimum hamming weight and then new viruses (new  vectors) are produced from the two classes of strong and common viruses. New codes are then generated by using new $\mathbf{x}$ vectors and their hamming weights are calculated. If the codes are self–dual and their hamming weights are at least 12 (this is the advantage of our search algorithm, we can fix the interval for the hamming weight), then the codes are saved. In the next step, the anti-virus phase, which is an immune-system process, is applied to the population to cope with uncontrolled growth of the viruses. If some termination criteria are met, the optimisation process is terminated. Next, the $[\gamma, \beta]$  values of the new generator matrices are checked and the previous generator matrix is replaced with one having the highest $[\gamma, \beta]$ parameters. Then the viruses are classified again and previous steps are repeated until the termination criteria of the optimisation algorithm are met. The details of the algorithm are given as a flow-chart in Figure~\ref{VOA}. In this algorithm, we set the population size to be 1000, the number of iteration to be 500 for each generator matrix whose neighbours we search for. We also set the termination criteria for the algorithm to be 50 runs, which means the algorithm will stop after searching for the possible neighbours of 50 different generator matrices.  The proposed algorithms are run on a workstation with Intel Xeon 4.0 GHz processor and 64 GByte RAM.  All the upcoming computational results were obtained by performing searches in the software package MAGMA (\cite{MAGMA}).

\newpage
\section{Generator Matrices}

In this section, we define generator matrices of the form $[I_{36} \ | \ \tau_6(v)]$ where $v \in M_6(\mathbb{F}_2)G,$ for  some groups of order $6$ and some $6 \times 6$ matrices that we later use as our initial generator matrices.

\begin{enumerate}
\item[I.] Let $\mathcal{C}_6$ be cyclic group of order 6 and $v_1 \in M_6(\mathbb{F}_2)\mathcal{C}_6.$
Then:
$$\tau_6(v)=\begin{pmatrix}
A&B\\
B'&A
\end{pmatrix}$$
where
$$A=CIRC(A_1,A_2,A_3),$$
$$B=CIRC(A_4,A_5,A_6),$$
$$B'=CIRC(A_6,A_4,A_5),$$
and where $A_i \in M_6(\mathbb{F}_2).$ Now we define two generator matrices of the following forms:
\begin{itemize}
\item[1.]
\begin{equation}
\mathcal{G}_1=[I_{36} \ | \ \tau_6(v_1)],
\end{equation}
with
$$A_1=circ(a_1,a_2, a_3, a_4, a_5, a_6),~~ A_2=circ(a_7,a_8,a_{9},a_{10},a_{11},a_{12}),$$
$$A_3=circ(a_{13},a_{14}, a_{15}, a_{16}, a_{17}, a_{18}),~~ A_4=circ(a_{19},a_{20},a_{21},a_{22},a_{23},a_{24}),$$
$$A_5=circ(a_{25},a_{26}, a_{27}, a_{28}, a_{29}, a_{30}),~~ A_6=circ(a_{31},a_{32},a_{33},a_{34},a_{35},a_{36}),$$
\item[2.]
\begin{equation}
\mathcal{G}_2=[I_{36} \ | \ \tau_6(v_1)],
\end{equation}
with
$$A_1=rcirc(a_1,a_2, a_3, a_4, a_5, a_6),~~ A_2=rcirc(a_7,a_8,a_{9},a_{10},a_{11},a_{12}),$$
$$A_3=rcirc(a_{13},a_{14}, a_{15}, a_{16}, a_{17}, a_{18}),~~ A_4=rcirc(a_{19},a_{20},a_{21},a_{22},a_{23},a_{24}),$$
$$A_5=rcirc(a_{25},a_{26}, a_{27}, a_{28}, a_{29}, a_{30}),~~ A_6=rcirc(a_{31},a_{32},a_{33},a_{34},a_{35},a_{36}),$$
\end{itemize}
\item[II.] Let $\mathcal{D}_6$ be dihedral group of order 6 and $v_2 \in M_6(\mathbb{F}_2)\mathcal{D}_6.$
Then:
$$\tau_6(v)=\begin{pmatrix}
A&B\\
B&A
\end{pmatrix}$$
where
$$A=CIRC(A_1,A_2,A_3),$$
$$B=RCIRC(A_4,A_5,A_6),$$
and where $A_i \in M_6(\mathbb{F}_2).$ Now we define two generator matrices of the following forms:
\begin{itemize}
\item[1.]
\begin{equation}
\mathcal{G}_3=[I_{36} \ | \ \tau_6(v_2)],
\end{equation}
with
$$A_1=circ(a_1,a_2, a_3, a_4, a_5, a_6),~~ A_2=circ(a_7,a_8,a_{9},a_{10},a_{11},a_{12}),$$
$$A_3=circ(a_{13},a_{14}, a_{15}, a_{16}, a_{17}, a_{18}),~~ A_4=circ(a_{19},a_{20},a_{21},a_{22},a_{23},a_{24}),$$
$$A_5=circ(a_{25},a_{26}, a_{27}, a_{28}, a_{29}, a_{30}),~~ A_6=circ(a_{31},a_{32},a_{33},a_{34},a_{35},a_{36}),$$
\item[2.]
\begin{equation}
\mathcal{G}_4=[I_{36} \ | \ \tau_6(v_2)],
\end{equation}
with
$$A_1=rcirc(a_1,a_2, a_3, a_4, a_5, a_6),~~ A_2=rcirc(a_7,a_8,a_{9},a_{10},a_{11},a_{12}),$$
$$A_3=rcirc(a_{13},a_{14}, a_{15}, a_{16}, a_{17}, a_{18}),~~ A_4=rcirc(a_{19},a_{20},a_{21},a_{22},a_{23},a_{24}),$$
$$A_5=rcirc(a_{25},a_{26}, a_{27}, a_{28}, a_{29}, a_{30}),~~ A_6=rcirc(a_{31},a_{32},a_{33},a_{34},a_{35},a_{36}),$$
\end{itemize}
\item[III.] Let $\mathcal{D}_6$ be dihedral group of order 6 and $v_3 \in M_6(\mathbb{F}_2)\mathcal{D}_6.$
Then:
$$\tau_6(v)=\begin{pmatrix}
A&B\\
B^T&A^T
\end{pmatrix}$$
where
$$A=CIRC(A_1,A_2,A_3),$$
$$B=CIRC(A_4,A_5,A_6),$$
and where $A_i \in M_6(\mathbb{F}_2).$ Now we define two generator matrices of the following forms:
\begin{itemize}
\item[1.]
\begin{equation}
\mathcal{G}_5=[I_{36} \ | \ \tau_6(v_3)],
\end{equation}
with
$$A_1=circ(a_1,a_2, a_3, a_4, a_5, a_6),~~ A_2=circ(a_7,a_8,a_{9},a_{10},a_{11},a_{12}),$$
$$A_3=circ(a_{13},a_{14}, a_{15}, a_{16}, a_{17}, a_{18}),~~ A_4=circ(a_{19},a_{20},a_{21},a_{22},a_{23},a_{24}),$$
$$A_5=circ(a_{25},a_{26}, a_{27}, a_{28}, a_{29}, a_{30}),~~ A_6=circ(a_{31},a_{32},a_{33},a_{34},a_{35},a_{36}),$$
\item[2.]
\begin{equation}
\mathcal{G}_6=[I_{36} \ | \ \tau_6(v_3)],
\end{equation}
with
$$A_1=rcirc(a_1,a_2, a_3, a_4, a_5, a_6),~~ A_2=rcirc(a_7,a_8,a_{9},a_{10},a_{11},a_{12}),$$
$$A_3=rcirc(a_{13},a_{14}, a_{15}, a_{16}, a_{17}, a_{18}),~~ A_4=rcirc(a_{19},a_{20},a_{21},a_{22},a_{23},a_{24}),$$
$$A_5=rcirc(a_{25},a_{26}, a_{27}, a_{28}, a_{29}, a_{30}),~~ A_6=rcirc(a_{31},a_{32},a_{33},a_{34},a_{35},a_{36}),$$
\end{itemize}
\end{enumerate}

We now take each generator matrix $\mathcal{G}_i$ and use them to search for binary $[72,36,12]$ self-dual codes which we later use as our initial generator matrices that correspond to the initial binary self-dual codes that we want to search for neighbours of. We list our finding in Table~\ref{genmat}.

\begin{table}[h!]
\caption{New Type I $[72,36,12]$ Codes from $\mathcal{G}_i$}
\resizebox{0.65\textwidth}{!}{\begin{minipage}{\textwidth}\label{genmat}
\centering
\begin{tabular}{ccccccccccc}
\hline
Generator Matrix   & $r_{A_1}$   &      $r_{A_2}$     & $r_{A_3}$           & $r_{A_4}$   &      $r_{A_5}$    & $r_{A_6}$  &   $\gamma$ & $\beta$ & $|Aut(C_i)|$ \\ \hline
$\mathcal{G}_1$ &  $(0,0,0,0,0,1)$ & $(0,0,0,0,0,1)$&   $(0,0,0,0,0,1)$&   $(0,0,0,0,0,1)$&   $(0,0,0,0,0,1)$& $(0,0,0,0,0,1)$& $0$      & $165$   & $72$         \\ \hline
$\mathcal{G}_2$ &  $(1,0,0,0,1,1)$ & $(0,0,0,0,0,1)$&   $(0,1,0,1,0,0)$&   $(0,1,0,0,1,1)$&   $(1,1,1,1,1,0)$& $(1,0,0,1,0,1)$& $0$      & $315$   & $72$         \\ \hline
$\mathcal{G}_3$ &  $(1, 1, 0, 1, 0, 0)$ & $(1, 0, 1, 1, 1, 1 )$&   $(1, 0, 0, 1, 1, 0)$&   $(1, 0, 0, 1, 1, 1)$&   $(0, 0, 1, 1, 0, 1 )$& $(0, 0, 0, 1, 1, 1)$& $0$      & $255$   & $36$         \\ \hline
$\mathcal{G}_4$ &  $(0, 1, 0, 1, 1, 1 )$ & $(1, 1, 0, 1, 1, 0)$&   $(1, 1, 1, 0, 1, 0 )$&   $(1, 0, 0, 1, 1,1)$&   $( 0, 0, 1, 0, 1, 0)$& $(1, 1, 0, 0, 1, 0)$& $0$      & $309$   & $72$         \\ \hline
$\mathcal{G}_5$ &  $(1, 1, 0, 1, 0, 0)$ & $(0, 0, 0, 1, 0, 1 )$&   $(1, 1, 0, 1, 0, 0)$&   $(1, 0, 1, 0, 0, 1 )$&   $(1, 1, 0, 0, 0, 1 )$& $(1, 1, 0, 0, 1, 0)$& $36$      & $537$   & $72$         \\ \hline
$\mathcal{G}_6$ &  $(1, 0, 0, 0, 0, 0 )$ & $(0, 0, 1, 1, 0, 1)$&   $(1, 0, 0, 1, 1, 1)$&   $(1, 1, 0, 0, 0, 1 )$&   $(0, 1, 1, 0, 0, 1 )$& $(1, 0, 0, 1, 1, 0 )$& $0$      & $231$   & $36$         \\ \hline
\end{tabular}
\end{minipage}}
\end{table}

We next search for possible neighbours of the binary codes listed in Table~\ref{genmat}. We obtain around 800 Type I neighbours for each code $\mathcal{G}_i$ from Table~\ref{genmat}, but we only present the codes whose parameters in their weight enumerators were not previously known.

\subsection{New Type I $[72,36,12]$ Codes by a Hybrid Search Technique from $\mathcal{G}_1$ }

We started with the generator matrix $\mathcal{G}_1$ given in Table \ref{genmat} and applied the search algorithm given in Figure~\ref{VOA}. We obtained 443 new Type I binary $[72,36,12]$ self-dual codes and their parameters are:

\begin{equation*}
\begin{array}{l}
(\gamma =0,\  \beta =\{ 172, 175, 176, 178, 181, 184, 188, 193, \}), \\[5pt]

(\gamma =1,\   \beta =\{179, 183,  189, 190, 194, 196, 198, 202, 204, 205   \}), \\[5pt]

(\gamma =2,\    \beta =\{185, 190, 194, 197, 198,  202, 207, 208, 212, 214, 215,  216, 217, 220, 236  \}), \\[5pt]

(\gamma =3,\    \beta =\{201, 205,  207, 209, 211, 213, 214, 223, 225, 226, 229    \}), \\[5pt]

(\gamma =4,\    \beta =\{209,  210, 212, 213, 214, 215, 216,  218, 219, 220, 221, 223, 224, 227, 228,  235, \\ 241, 242, 243, 247, 257,  \}), \\[5pt]

(\gamma =5,\     \beta =\{ 219, 223, 224, 225, 226, 227, 230, 232, 233,  237, 238, 239, 240, 241, 244, 246,\\  247, 248, 250, 251, 252, 253    \}) \\[5pt]

(\gamma =6,\     \beta =\{ 224, 226, 228, 234, 235, 239, 240, 241, 244,  245, 247, 249, 250, 251,  254,  \}) \\[5pt]

(\gamma =7,\     \beta =\{ 230, 231, 232, 233, 234, 236, 237, 238, 239, 240, 241, 242, 243, 244, 245, 246,\\
247, 248, 249, 250, 251, 252, 253, 254, 255, 256, 257, 258, 259, 261, 263, 264, 269, 271, 272,\\
 273, 275, 276, \}) \\[5pt]

(\gamma =8,\     \beta =\{236, 238, 239, 240, 241, 242, 243, 244, 245, 246, 247, 248, 249, 250, 251, 252,\\
253, 254, 255, 256, 257, 258, 259, 260, 261, 262, 263, 264, 265, 266, 267, 268, 269, 270, 271,\\
 272, 273, 274,  276,  280, 281    \}) \\[5pt]

(\gamma =9,\     \beta =\{ 234,  240, 241, 242, 244, 245, 247, 248, 249, 250, 251, 254, 256, 257, 259, 262,\\
 263, 265, 266, 268, 269, 271, 272, 274, 275, 276, 277, 278, 280, 281, 290, 295,  \}) \\[5pt]

(\gamma =10,\     \beta =\{234, 236, 238, 240, 242, 243, 244, 245, 246, 248, 250, 252, 253, 254, 255, 256, \\
257, 258,  259, 260, 261, 262, 263, 264, 265, 266, 267, 268,  270, 271, 272, 273, 274, 275, 276,\\
277,278, 279, 280, 281, 282, 283, 284, 285, 286, 287, 288, 290, 293, 295, \}) \\[5pt]

(\gamma =11,\     \beta =\{240, 241, 243, 244, 245, 246, 247, 248, 249, 250, 251, 252, 253, 254, 255, 256,\\
 257, 258, 259, 260, 261, 262, 263, 264, 265, 266, 267, 268, 269, 270, 271, 272, 273, 274, 275, 276,\\
 277, 278, 279,  280,  281, 282, 283, 284, 285, 286, 287, 288, 289, 290, 291, 292, 296, 297, 300,\}) \\[5pt]

(\gamma =12,\     \beta =\{ 243, 246, 250, 251,  254, 255, 256, 257, 258, 259, 260, 261,  262, 263, 265, 266,\\
 268, 271,  273, 274, 275, 276, 277, 280, 281, 283, 284,  286, 287, 292, 293, 294, 295, 297, 300   \}) \\[5pt]

\end{array}%
\end{equation*}
\begin{equation*}
\begin{array}{l}

(\gamma =13,\     \beta =\{ 251, 253, 256, 257, 258, 260, 262, 263, 264, 266, 267, 268, 269, 270, 271, 272,\\
 273, 274, 275, 276, 277, 278, 279, 280, 281, 282, 283, 284, 285, 286, 287, 288, 289, 290, 291, 292,\\
  293, 294, 295, 296, 297, 298, 302, 303, 307, 310, 312, \}) \\[5pt]

(\gamma =14,\     \beta =\{263, 265, 266,  268, 269, 270, 272,  274, 277, 278, 279, 280, 281, 282,  283,\\
 284, 285, 286, 287, 289, 290, 291, 293, 296, 297, 302, 304, 308, 314,  \}) \\[5pt]

(\gamma =15,\     \beta =\{268,  272, 277, 278, 280, 283, 287, 291, 295, 297,   \}) \\[5pt]

(\gamma =16,\     \beta =\{286, 292, 295, 300, 304  \}) \\[5pt]

\end{array}%
\end{equation*}

\subsection{ New Type I $[72,36,12]$ Codes by a Hybrid Search Technique from $\mathcal{G}_2$ }

We started with the generator matrix $\mathcal{G}_2$ given in Table \ref{genmat} and applied the search algorithm given in Figure~\ref{VOA}. We obtained 153 new Type I binary $[72,36,12]$ self-dual codes and their parameters are:

\begin{equation*}
\begin{array}{l}

(\gamma =0,\  \beta =\{241,  268, 274, 287  \}), \\[5pt]

(\gamma =1,\   \beta =\{ 214, 228, 230, 231, 241, 251, 265, 267, 268, 269, 273, 278,  \}), \\[5pt]

(\gamma =2,\   \beta =\{ 221, 234, 235,  237, 238, 252, 253, 259, 265, 272, 274, 282, 283, 292  \}), \\[5pt]

(\gamma =3,\   \beta =\{224, 227, 230, 234, 237 242, 246, 253, 259, 271 \}), \\[5pt]

(\gamma =4,\   \beta =\{222, 233, 238, 239, 244, 246, 248, 251,  258, 268  \}), \\[5pt]

(\gamma =5,\   \beta =\{243, 245, 254, 257  \}), \\[5pt]

(\gamma =6,\   \beta =\{259  \}), \\[5pt]

(\gamma =7,\   \beta =\{260, 265, 267, 270   \}), \\[5pt]

(\gamma =8,\   \beta =\{275, 278, 285   \}), \\[5pt]

(\gamma =9,\   \beta =\{286, 293   \}), \\[5pt]

(\gamma =10,\   \beta =\{230, 235, 291   \}), \\[5pt]

(\gamma =11,\   \beta =\{237, 239, 294, 295   \}), \\[5pt]

(\gamma =12,\   \beta =\{234, 241, 298  \}), \\[5pt]

(\gamma =13,\   \beta =\{245, 259, 261, 265, 299, 304   \}), \\[5pt]

(\gamma =14,\   \beta =\{253, 255, 256, 257, 264, 271, 273, 275, 276, 288, 292, 294, 295, 298, 299,\\ 300, 301, 305, 306, 307, 309   \}), \\[5pt]
\end{array}%
\end{equation*}
\begin{equation*}
\begin{array}{l}

(\gamma =15,\   \beta =\{269, 271, 273, 275, 276, 281, 282, 284, 285, 286, 288, 289, 290, 292, 293,\\ 294, 298, 299, 300, 301, 302, 303, 305, 306, 310, 314, 319   \}), \\[5pt]

(\gamma =16,\   \beta =\{285, 288, 296, 297, 298, 299, 301, 303, 305, 307, 308, 310, 321  \}), \\[5pt]

(\gamma =17,\   \beta =\{283, 287, 294, 296, 304, 307, 308, 314, 317, 319, 327   \}), \\[5pt]

(\gamma =18,\   \beta =\{307, 313   \}), \\

\end{array}%
\end{equation*}

\subsection{New Type I $[72,36,12]$ Codes by a Hybrid Search Technique from $\mathcal{G}_3$ }

We started with the generator matrix $\mathcal{G}_3$ given in Table \ref{genmat} and applied the search algorithm given in Figure~\ref{VOA}. We obtained 71 new Type I binary $[72,36,12]$ self-dual codes and their parameters are:

\begin{equation*}
\begin{array}{l}

(\gamma =0,\  \beta =\{229 \}), \\[5pt]

(\gamma =1,\   \beta =\{201, 224  \}), \\[5pt]

(\gamma =2,\   \beta =\{203, 204, 213, 218, 230   \}), \\[5pt]

(\gamma =3,\   \beta =\{228, 233, 239, 247 \}), \\[5pt]

(\gamma =4,\   \beta =\{211, 250, 252 \}), \\[5pt]

(\gamma =5,\   \beta =\{256, 258, 260, 262, 270  \}), \\[5pt]

(\gamma =6,\   \beta =\{262, 268, 271, 278  \}), \\[5pt]

(\gamma =7,\   \beta =\{266, 268, 274, 279, 283, 284, 285   \}), \\[5pt]

(\gamma =8,\   \beta =\{279, 284, 287, 288, 289, 290  \}), \\[5pt]

(\gamma =9,\   \beta =\{283, 284, 285, 299  \}), \\[5pt]

(\gamma =10,\   \beta =\{224, 292, 294, 296, 297, 298, 299, 300 \}), \\[5pt]

(\gamma =11,\   \beta =\{293, 298, 299, 303, 307, 308, 310  \}), \\[5pt]

(\gamma =12,\   \beta =\{299, 305, 307  \}), \\[5pt]

(\gamma =13,\   \beta =\{248, 301, 305, 313, 316, 318, 323, 325  \}), \\[5pt]

(\gamma =14,\   \beta =\{310, 318  \}), \\[5pt]

(\gamma =15,\   \beta =\{309, 312 \}), \\[5pt]

\end{array}%
\end{equation*}

 \subsection{ New Type I $[72,36,12]$ Codes by a Hybrid Search Technique from $\mathcal{G}_4$ }

We started with the generator matrix $\mathcal{G}_4$ given in Table \ref{genmat} and applied the search algorithm given in Figure~\ref{VOA}. We obtained 77 new Type I binary $[72,36,12]$ self-dual codes and their parameters are:

\begin{equation*}
\begin{array}{l}

(\gamma =0,\  \beta =\{250, 266\}), \\[5pt]

(\gamma =1,\   \beta =\{263, 271\}), \\[5pt]

(\gamma =2,\   \beta =\{227, 246, 247, 263, 271\}), \\[5pt]

(\gamma =3,\   \beta =\{264, 265, 279,  295\}), \\[5pt]

(\gamma =4,\   \beta =\{240, 254, 256, 261, 267, 270, 272, 276\}), \\[5pt]

(\gamma =5,\   \beta =\{267\}), \\[5pt]

(\gamma =7,\   \beta =\{277\}), \\[5pt]

(\gamma =12,\   \beta =\{238\}), \\[5pt]

(\gamma =13,\   \beta =\{247, 250, 252, 254\}), \\[5pt]

(\gamma =14,\   \beta =\{258, 259, 261, 262, 303, 312, 313, 316, 322\}), \\[5pt]

(\gamma =15,\   \beta =\{261, 264, 274, 279, 296, 308, 311, 315, 316\}), \\[5pt]

(\gamma =16,\   \beta =\{281, 287, 290, 291, 294, 302, 306, 309, 311, 313, 315, 324\}), \\[5pt]

(\gamma =17,\   \beta =\{286, 291, 292, 295, 300, 301, 302, 306, 310, 311, 312 , 315, 316, 320\}), \\[5pt]

(\gamma =18,\   \beta =\{304, 311, 323\}), \\[5pt]

(\gamma =19,\   \beta =\{314, 319\}). \\[5pt]
\end{array}%
\end{equation*}

\subsection{ New Type I $[72,36,12]$ Codes by a Hybrid Search Technique from $\mathcal{G}_5$ }

We started with the generator matrix $\mathcal{G}_5$ given in Table \ref{genmat} and applied the search algorithm given in Figure~\ref{VOA}. We obtained 727 new Type I binary $[72,36,12]$ self-dual codes and their parameters are:

\begin{equation*}
\begin{array}{l}

(\gamma =12,\   \beta =\{329, 333, 340, 341  \}), \\[5pt]

(\gamma =13,\   \beta =\{349  \}), \\[5pt]

(\gamma =14,\   \beta =\{333, 336, 337, 338, 339, 341, 342, 345, 347, 349, 353, 354, 355  \}), \\[5pt]

(\gamma =15,\   \beta =\{332, 352  \}), \\[5pt]

(\gamma =16,\   \beta =\{327, 328, 332, 333, 335, 337, 338, 339, 340, 341, 343, 344, 345, 346, 347, 348,\\
 349, 350, 351, 352, 353, 355, 356, 357, 358, 359, 360, 361, 362, 363, 364, 365, 366, 367, 368, \\
 369, 371, 372, 374, 375, 376, 378, 379 \}), \\[5pt]

(\gamma =17,\   \beta =\{339, 356, 357, 360, 361, 363, 364, 366, 367, 368, 369, 371, 372, 373\}), \\[5pt]

(\gamma =18,\   \beta =\{332, 337, 343, 344, 346, 347, 349,  352, 353, 355, 356, 358, 361, 362, 365, 367,\\
370, 371, 373, 374, 375, 376, 377,  379, 380, 382, 383, 385, 386, 388, 389, 392, 394\}), \\[5pt]

(\gamma =19,\   \beta =\{345, 348, 350, 354, 356, 358, 359, 360, 361, 362, 363, 364, 365, 366, 367, 368, \\
369, 370, 371, 372, 373, 374, 375, 376, 377, 379, 381, 382, 383, 385, 386, 388, 389, 390, 393, \\
395, 397, 398\}), \\[5pt]

(\gamma =20,\   \beta =\{334, 340, 341, 342, 345, 346, 347, 348, 349, 350, 351, 352, 353, 354, 355,
\\356, 357, 358,  359, 360, 361, 362, 363, 364, 365, 366, 367, 368, 369, 370, 371, 372, 373, 374,
\\ 375, 376, 377, 378, 379, 380, 381, 382, 383, 384, 385, 386, 387, 388, 389, 390, 392, 391, 393,
\\ 394, 395, 396, 397, 398, 399, 400, 401, 402, 403, 404, 405, 406, 407, 410, 411, 412, 426 \}), \\[5pt]

(\gamma =21,\   \beta =\{357, 358, 359, 361, 363, 364, 365, 366, 367, 368, 369, 370, 371, 372, 373,
\\  374, 375, 376, 377, 378, 379, 380, 381, 382, 383, 384, 385, 386, 387, 388, 389, 390, 391,
\\ 392, 393, 394, 395, 396, 397, 398, 399, 400, 401, 402, 403, 404, 405, 406, 407, 408,
\\  409, 410, 411, 413, 415\}), \\[5pt]

(\gamma =22,\   \beta =\{350, 352, 354, 355, 357, 358, 359, 360, 361, 362, 363, 364, 365, 366,
\\ 367, 368, 369, 370, 371, 372, 373, 374, 375, 376, 377, 378, 379, 380, 381, 382, 383, 384,
\\ 385, 386, 387, 388, 389, 390, 391, 392, 393, 394, 395, 396, 397, 398, 399, 400, 401, 402, 403,
\\ 404 405, 406, 407, 408, 409, 410, 411, 412, 413, 414, 415, 416, 417, 418, 419, 420, 422, 423,
\\ 425, 429, 435, 437\}), \\[5pt]

(\gamma =23,\   \beta =\{361, 366, 367, 370, 371, 372, 373, 374, 375, 378, 379, 380, 381, 382, 383,
\\ 384, 385, 386, 387, 388, 389, 390, 391, 392, 393, 394, 395, 396, 397, 398, 399, 400, 401, 402,
\\ 403, 404, 405, 406, 407, 408, 409, 410, 412, 413, 414, 415, 416, 417, 418, 419, 420, 421, 422,
\\  424, 425, 426, 428, 431, 433\}), \\[5pt]
\end{array}%
\end{equation*}

\begin{equation*}
\begin{array}{l}
(\gamma =24,\   \beta =\{360, 365, 367, 369, 371, 372, 373, 374, 376, 377, 378, 379, 380, 382, 384, 385,
\\ 386, 388, 389, 390, 391, 394, 395, 396, 397, 398, 399, 400, 401, 403,  405, 406, 407, 409, 410, 412,\\
413, 414, 415, 416, 417, 418, 419, 420, 421, 422, 423, 424, 425, 426, 428, 431, 432, 434, 436,  437,\\
 438, 440,  443, 447 \}), \\[5pt]

(\gamma =25,\   \beta =\{378, 381, 382, 383, 384, 385, 386, 387, 388, 389, 390, 391, 392, 393, 394, 395, 396,
\\ 397, 398, 399, 400, 401, 402, 403, 404, 405, 406, 407, 408, 409, 410, 411, 412, 413, 414, 415, 416,\\
 417, 418, 419, 420, 421, 422, 423, 424, 425, 426, 427, 428, 429, 430, 431, 432, 433, 434, 435, 437,\\
  438, 439, 444, 447 \}), \\[5pt]

(\gamma =26,\   \beta =\{378, 381, 382, 383, 384, 385, 386, 387, 388, 389, 390, 391, 392, 393, 394, 395,
\\ 396, 397, 398, 399, 400, 401, 402, 403, 404, 405, 406, 407, 408, 409, 410, 411, 412, 413, 414, 415,
\\ 416, 417, 418, 419, 420, 421, 422, 423, 424, 425, 426, 427, 428, 429, 430, 431, 432, 433, 434, 435,
\\ 436, 437, 438, 439, 440, 442, 443, 444, 446, 447, 448, 449, 450 \}), \\[5pt]

(\gamma =27,\   \beta =\{390, 396, 397, 398, 409, 410, 411, 412, 413, 414, 415, 416, 417, 418, 419,
\\ 420, 421, 422, 424, 425, 427, 428, 429, 430, 431, 432, 433, 434, 435, 436, 437, 438, 439, 440,
\\ 441, 442, 443, 445, 446, 447, 448, 449\}), \\[5pt]

(\gamma =28,\   \beta =\{391, 394, 397, 401, 402, 404, 406, 408, 409, 410, 412, 413, 415, 417, 418,
\\ 419, 420, 421, 422, 423, 424, 425, 426, 427, 428, 429, 430, 431, 432, 434, 435, 436, 437, 438,
\\ 439, 441, 442, 443, 445, 446, 448, 449, 451, 456, 462, 471, 478\}), \\[5pt]

(\gamma =29,\   \beta =\{411, 412, 417, 420, 423, 424, 427, 429, 431, 432, 436, 439, 440, 441, 442,
\\ 443, 444, 445, 447, 449, 450, 452, 455, 456, 457, 460, 461, 463, 464, 467, 469, 472 \}), \\[5pt]

(\gamma =30,\   \beta =\{436, 437, 443, 446, 450, 451, 454, 456, 457, 461 \}), \\[5pt]

(\gamma =31,\   \beta =\{434, 436, 440, 445, 448, 450, 463, 468, 469, 472 \}), \\[5pt]

(\gamma =32,\   \beta =\{451, 463, 464\}), \\[5pt]

\end{array}%
\end{equation*}

\subsection{New Type I $[72,36,12]$ Codes by a Hybrid Search Technique from $\mathcal{G}_6$ }

We started with the generator matrix $\mathcal{G}_6$ given in Table \ref{genmat} and applied the search algorithm given in Figure~\ref{VOA}. We obtained 39 new Type I binary $[72,36,12]$ self-dual codes and their parameters are:

\begin{equation*}
\begin{array}{l}

(\gamma =0,\  \beta =\{191\}), \\[5pt]

(\gamma =1,\   \beta =\{ 221\}), \\[5pt]

(\gamma =2,\   \beta =\{ 196, 210, 211\}), \\[5pt]

(\gamma =5,\   \beta =\{ 263\}), \\[5pt]

(\gamma =6,\   \beta =\{194\}), \\[5pt]

(\gamma =11,\   \beta =\{ 234\}), \\[5pt]

(\gamma =12,\   \beta =\{ 244, 245, 304\}), \\[5pt]

(\gamma =13,\   \beta =\{240, 243, 249, 255, 306, 308, 309\}), \\[5pt]

(\gamma =15,\   \beta =\{259, 262, 270, 307, 313, 318, 328\}), \\[5pt]

(\gamma =16,\   \beta =\{ 279, 284, 289, 293, 322, 323, 325\}), \\[5pt]

(\gamma =17,\   \beta =\{ 273, 289, 303, 316, 322\}), \\[5pt]

(\gamma =18,\   \beta =\{ 301, 306\}). \\[5pt]

\end{array}%
\end{equation*}

\section{Conclusion}
In this work, we presented a hybrid search algorithm  based on the VOA for the problem of finding neighbours of binary self-dual codes by using a number of generator matrices of the form $[I_{36} \ | \ \tau_6(v)]$.
As a result, we constructed 1471 new binary $[72, 36, 12]$ self-dual codes.
A suggestion for future work is to consider generator matrices of the form $[I_{kn} \ | \ \tau_k(v)]$ for groups of orders different than 6 and for values of $k$ different than 6, to search for extremal binary self-dual codes of different lengths. Another suggestion is to consider generator matrices of the form $[I_{kn} \ | \ \tau_k(v)]$ over different alphabets,  and explore the binary images of the codes under the Gray maps.

\end{document}